\documentclass[12pt]{article}

\usepackage{amsmath,amsthm,amssymb}
\usepackage{answers}
\usepackage{setspace}
\usepackage{graphicx}
\usepackage{enumitem}
\usepackage{multicol}
\usepackage{mathrsfs}
\usepackage{appendix}
\usepackage[margin=2cm]{geometry} 
\usepackage{floatrow}
\usepackage{mathtools}  
\usepackage{commath}    
\usepackage{listings}
\usepackage{subfig}
\usepackage{cite}

\usepackage{authblk}

\title{Can bus bunching reduce waiting time?}
\begin{document}
\setcounter{Maxaffil}{0}
\renewcommand\Affilfont{\itshape\small}
\author[1, 2]{Luca Vismara}
\author[2]{Vee-Liem Saw}
\author[2, *]{Lock Yue Chew}
\affil[1]{Interdisciplinary Graduate Programme, 61 Nanyang Drive, Nanyang Technological University, Singapore 637335}
\affil[2]{Division of Physics and Applied Physics, School of Physical and Mathematical Sciences, 21 Nanyang Link, Nanyang Technological University, Singapore 637371}
\affil[*]{Corresponding author: Lock Yue Chew, lockyue@ntu.edu.sg}

\maketitle

\begin{abstract}
Bus bunching is ostensibly regarded as a detrimental phenomenon in bus systems. We study a bus loop with two bus stops, one regular and one spike bus stop, where bunched buses can outperform staggered buses. The spike bus stop models a bus stop connected to a train or metro service, where passengers arrive in groups at periodic intervals (spikes). We introduce the configuration of synchronised bunched buses, where bunched buses wait for the spike in demand. For a wide range of parameters, synchronised bunched buses outperform perfectly staggered buses in terms of minimising the waiting time of commuters. We present an analytical formulation of the average waiting time in the case of bunched buses, synchronised bunched buses and perfectly staggered buses with the consideration of different passenger demand, number of buses, and bus capacity. We demonstrate the validity of our analytical results through an agent-based simulation of the bus loop system.
\end{abstract}

\section{Introduction}
Bus bunching is generally regarded as undesirable in bus systems \cite{Daganzo2009,Saw2019intelligent,Chew2019,Abkowitz1984,rossetti1998,Hickman2001,Fu2002,cats2011,bartholdi2012,Moreira-Matias2016,Quek2020,Saw2019,Wang2020}. As explained in Refs. \cite{Vismara2021conf, Azfar2018, Osuna1972}, waiting time for passengers is reduced by a staggered bus configuration when passengers arrive continuously at bus stops. In the transportation literature, the common assumption regarding passenger arrival is that the number of passengers arriving at a bus stop grows continuously and linearly with time \cite{Osuna1972, Saw2019, Saw2019intelligent, Saw2019no, Saw2020chaos, Saw2020, Saw2021, Vismara2021, Vismara2021conf, Daganzo2009, Hickman2001, Fu2002, Fu2003, Newell1974, cats2011}, either deterministically or stochastically. Here we want to study the case of a system with a \emph{spike bus stop}.

Passengers arrive at a \emph{spike bus stop} in batches of $p$ at a time at periodic intervals of $T_s$ unit times.
A real-world example of a bus stop having this property is a bus stop connected to a rapid transit system or a  train station with periodic service. Passengers waiting at the bus stop are still increasing with time, as in the cases examined by the literature, but they increase periodically in groups of $p$, rather than one by one or continuously. To the knowledge of the authors, this kind of bus stop has not been studied before in the literature.
Allowing $p$ to be stochastic makes the model more realistic, however, for analytical tractability and to give physical insights, we consider $p$ as a constant in this paper.

In this work, we study the case of a bus loop with one \emph{spike bus stop} and one \emph{regular bus stop}, as defined in section \ref{sec:3unlimited}. The \emph{regular bus stop} is analogous to the bus stops examined in Refs. \cite{Saw2019intelligent, Vismara2021, Vismara2021conf}. Through analytical models, we compute the average waiting time for three configurations: bunched buses, synchronised bunched buses (as defined in section \ref{ssec:B}) and fully staggered buses. The calculations are performed for the case of unlimited bus capacity in section \ref{sec:3unlimited} and by explicitly considering buses with limited capacity in section \ref{sec:limited}. We compare the three bus configurations in section \ref{sec:comparison} and we show that bunched buses synchronising with the spike in passenger demand can outperform perfectly staggered buses in the regime of high demand in the \emph{spike bus stop} and when bus capacity is limited. We also briefly generalise the results for a loop with multiple \emph{regular bus stops} in section \ref{ssec:generalisation}. In section \ref{ssec:simulation} we validate the theoretical results with a simulation. In section \ref{sec:conclusion3}, we summarise the results obtained and draw the conclusion and final remarks.

\section{Three scenarios with a \emph{spike bus stop}}\label{sec:3unlimited}

The system we set to study is a bus loop with two bus stops. At the first bus stop, passengers arrive at a constant rate of $s$ per unit time.
The second bus stop is called \emph{spike bus stop} and passengers arrive in spikes of $p$ people every $T_s$ unit times.
This bus loop is inspired by the ``Campus Rider'' in Nanyang Technological University that serves both regular bus stops around the campus and a bus stop connected to the ``Mass Rapid Transit'' line (MRT).
The first kind of bus stop, which we call \emph{regular bus stop}, is modelled in the conventional way \cite{Osuna1972, Saw2019, Saw2019intelligent, Saw2019no, Saw2020chaos, Saw2020, Saw2021, Vismara2021, Vismara2021conf, Daganzo2009, Hickman2001, Fu2002, Fu2003, Newell1974}. Passengers arrive at a constant rate $s$: defining $\Delta_t$ as the time elapsed since that bus stop was lastly served by a bus, the number of passengers waiting is $s \times \Delta_t$.
While it is more realistic to model passenger arrival as a stochastic process \cite{Hickman2001,Fu2002,cats2011,Moreira-Matias2016,Quek2020,Saw2019no}, however, for analytical tractability, in this work we consider deterministic passenger arrivals at a constant rate $s$.
In the analysis presented in this paper, we fix the rate at which passengers board as $l$ passengers per unit time and we assume passengers alight instantaneously at the opposite bus stop, not affecting the dwelling time. To simplify the expressions, it is convenient to work with the quantities $k = s / l$ and $P = p / l$ to represent passengers arriving at bus stops.
Without loss of generality, we set $l$ as one passenger per unit time (which effectively defines the unit of time) and refer to $k = s / l$ as the arrival constant of the \emph{regular bus stop} and to $P$ as the number of passengers at the \emph{spike bus stop}.
In this section, the capacity of the buses is unconstrained, while section \ref{sec:limited} explicitly considers the implications of limited capacity. All of the buses move at the same speed and take time $T$ to complete the loop without counting the dwell time at bus stops. If more than a bus is at the same bus stop, such as in a case of bus bunching, the load of passengers is equally shared, effectively multiplying the boarding rate. Notation wise, we use the square symbol $\square$ to refer to the \emph{regular bus stop} and the triangle symbol $\triangle$ for the \emph{spike bus stop}.
We measure distances in units of time, as the speed of the buses is fixed.

The following three sections explore the three configurations of buses tested in the bus loop defined above:
\begin{enumerate}[label=\Alph*]
    \item Bunched buses;
    \item \emph{Synchronised bunching}; \label{enum:sync}
    \item Perfectly staggered buses.
\end{enumerate}
Case \ref{enum:sync}, ``Synchronised bunching'', is a novel approach that aims at being easier to implement compared with dynamic control techniques, exploiting the periodicity of the arrivals at the \emph{spike bus stop}. The definition of this tactic is in section \ref{ssec:B}.

In the following sections, we calculate analytically the average waiting time for passengers at bus stops as a function of the crowdedness of the \emph{regular bus stop} $k=s/l$, the number of people arriving at the \emph{spike bus stop} $P=p/l$ and their period of arrival $T_s$ for $N$ buses.
We consider spikes with a relatively long period $T_s$ such that at most one spike occurs during a revolution.

\subsection{Bunched buses}\label{ssec:A}
It is well known that uncontrolled buses bunch \cite{Newell1974, Chew2019, Saw2019, Vismara2021conf} so this is a natural baseline and an important case to consider as bus bunching is not an uncommon occurrence in real-world bus systems.

The general idea to compute the average waiting time is to separate the contribution of the \emph{regular} $\square$ and \emph{spike} $\triangle$ bus stop.

For the \emph{regular bus stop}, using the fact that passengers board in a FIFO way (the first to arrive is the first to board), the average waiting time at the \emph{regular bus stop} is half the waiting time of the passenger who waited for the longest, which is the first to arrive at the bus stop after the last platoon of buses left the bus stop in the previous revolution. Passengers arrive and board in a linear fashion, and the last passenger to board is the one that arrives just before the platoon of buses leave, giving a waiting time of zero, hence the average is half the longest waiting time. The same idea is also used to compute the average waiting time in Ref. \cite{Vismara2021}. The waiting time at that bus stop is:
\begin{equation}\label{eq:w_s_A}
    \text{W}^\square_A = \frac{1}{2} \left( \bar{T}_A - \tau^\square_A \right) = \frac{\bar{T}_A}{2} \left(1 - \frac{k}{N}\right).
\end{equation}
The quantity $\bar{T}_A$ represents the effective time taken for the buses to complete the loop, including dwelling time. It is defined in Eq. \eqref{eq:tbar_A}. The time spent dwelling at the bus stop is $\tau^\square_A$ and it is proportional to the total passengers arrived, hence proportional to the time it takes to complete a revolution $\bar{T}_A$, times the arrival constant $k = s / l$. Having $N$ buses bunching, the dwelling time is reduced by a factor $1 / N$ as passengers can board in parallel the $N$ buses speeding up the process.
To compute the waiting time for the whole bus loop, the waiting time at each bus stop has to be weighted by the passengers boarding. To keep a consistent notation with $k$, we use the number of passengers boarding per revolution divided by the constant boarding rate $l$.
\begin{equation}\label{eq:p_s_A}
    \text{Ppl}^\square_A = k \bar{T}_A.
\end{equation}

In the case of the \emph{spike bus stop}, passengers arrive in a group of $P=p / l$. They all wait for the buses to arrive. At every revolution, the waiting time will be different, unless $T_s$ and $\bar{T}_A$ happen to be exactly in resonance (in this work we do not cover this possibility), so it is a reasonable assumption to consider an average value.
Assuming that the buses are already in a bunched state, and 
for analysis purpose we have the platoon of buses distributed uniformly around the bus loop when the spike happens with the arrival of the passengers at the \emph{spike bus stop}, the average time until the buses reach the bus stop is $\delta^\triangle = \bar{T}_A / 2$. On top of $\delta^\triangle$ time to wait, passengers need to board. The first passenger to board does not need to wait any longer, but the last passenger to board has to wait $P/N$ units of time extra, as boarding is conducted in parallel with $N$ buses, so on average the extra waiting is $1/2 \times P/N$:
\begin{equation}\label{eq:w_t_A}
    \text{W}^\triangle_A = \delta^\triangle + \frac{P}{2 N} = \frac{\bar{T}_A}{2} + \frac{P}{2 N}.
\end{equation}
Considering that passengers arrive at the \emph{spike bus stop} every $T_s$, the average number of passengers boarding from this bus stop over a revolution is
\begin{equation}\label{eq:p_t_A}
    \text{Ppl}^\triangle_A = P \frac{\bar{T}_A}{T_s}.
\end{equation}

The final equation needed is for $\bar{T}_A$, the average time taken to complete a revolution, which comprises of the time $T$ needed to drive, and the time spent at the bus stops boarding passengers $\tau^\square_A$ and $\tau^\triangle_A$:
\begin{equation}\label{eq:tbar_A}
    \bar{T}_A = T + \tau^\triangle_A + \tau^\square_A = T + \frac{P \bar{T}_A}{N T_s} + \frac{k \bar{T}_A}{N} = \frac{T}{1 - \frac{P}{N T_s} - \frac{k}{N}}.
\end{equation}
From the denominator of Eq. \eqref{eq:tbar_A} we see the two conditions of maximum crowdedness at which the bunched buses get stuck at one or the other bus stop. For the \emph{regular bus stop}, $N$ buses can board $N$ passengers per unit time, hence if passengers arrive at a rate $\ge N$ per unit time ($k \ge N$), the buses cannot finish boarding and $\bar{T}_A \to \infty$. For the \emph{spike bus stop}, as buses require $P / N$ unit times to board all the passengers from the spike, if $P / N \ge T_s$ a new spike arrives before the \emph{spike bus stop} is cleared, blocking the buses and causing $\bar{T}_A$ to diverge.
The average waiting time in the bus loop is computed combining Eqs. \eqref{eq:w_s_A} \eqref{eq:p_s_A} \eqref{eq:w_t_A} \eqref{eq:p_t_A} \eqref{eq:tbar_A}:
\begin{equation}\label{eq:w_A}
    \text{W}_A = \frac{\text{Ppl}^\triangle_A \times \text{W}^\triangle_A + \text{Ppl}^\square_A \times \text{W}^\square_A}{\text{Ppl}^\triangle_A + \text{Ppl}^\square_A} =\frac{P^2 + N \bar{T}_A \left(  P  +
    k \left(1 - \frac{k}{N} \right) \right) T_s }{2 N \left( P + k T_s\right)}.
\end{equation}

\subsection{Synchronised bunched buses}\label{ssec:B}
In this section, we introduce a novel approach to bus control. As dynamic control often requires specific infrastructure, complex algorithms and nonlocal information such as the position of other buses \cite{Daganzo2009,Saw2019intelligent,Chew2019,Abkowitz1984,rossetti1998,Hickman2001,Fu2002,cats2011,bartholdi2012,Moreira-Matias2016,Quek2020,Saw2019,Wang2020, Newell1974,Barnett1974, Alesiani2018, Fu2003,LEIVA20101186,Yavuz2010,Chiraphadhanakul2013,chen2015design,larrain2015generation,soto2017new}, we propose a much simpler yet effective technique: hold the platoon of bunched buses at the \emph{spike bus stop} until the spike happens. It is well known that buses tend to bunch, but bunched buses have a distinct advantage: they can distribute the load of passengers among them, speeding up boarding and reducing dwelling time. Another advantage is that $N$ buses can accommodate $N$ times as many passengers, although the problem of limited bus capacity is analysed in section \ref{ssec:B_lim}.
This technique, which we call \emph{synchronised bunching}, synchronises the effective time taken by buses to complete a loop (including dwelling time) $\bar{T}_B$ with the period of the spike $T_s$.

To calculate the waiting time in this scenario, we compute the waiting time at each bus stop and weight them by the number of passengers boarded from there. With the \emph{regular bus stop} the waiting time is computed in the same way as in Eq. \eqref{eq:w_s_A}, with the added condition that $\bar{T}_B = T_s$:
\begin{equation}\label{eq:w_s_B}
    \text{W}^\square_B = \frac{1}{2} \left(T_s - \tau_B^\square \right) = \frac{T_s}{2} \left( 1 - \frac{k}{N} \right).
\end{equation}
Following the same idea as in Eq. \eqref{eq:p_s_A},
\begin{equation}\label{eq:p_s_B}
    \text{Ppl}^\square_B = k  T_s.
\end{equation}

The waiting time at the \emph{spike bus stop} is very simple, as the passengers can start boarding immediately because the platoon of buses is waiting for the spike, so the only time to wait is the time it takes to board, which is equivalent to the second term in Eq. \eqref{eq:w_t_A} as boarding happens in parallel for bunched buses.
\begin{equation}\label{eq:w_t_B}
    \text{W}^\triangle_B = \frac{P}{2 N}.
\end{equation}
Given that the buses effectively take $T_s$ time to complete a revolution, the number of passengers boarding per revolution is $P$.
\begin{equation}\label{eq:p_t_B}
    \text{Ppl}^\triangle_B = P.
\end{equation}

Combining the equations above, the average waiting time for passengers of the loop when the \emph{synchronised bunching} technique is employed is
\begin{equation}\label{eq:w_B}
    \text{W}_B = \frac{\text{Ppl}^\triangle_B \times \text{W}^\triangle_B + \text{Ppl}^\square_B \times \text{W}^\square_B}{\text{Ppl}^\triangle_B + \text{Ppl}^\square_B} = \frac{P^2 + k {T_s}^2\left( N - k \right)}{2 N \left( P + k T_s \right)}.
\end{equation}

\subsection{Perfectly staggered buses}\label{ssec:C}
The calculations in this section are performed under the assumption of perfectly staggered buses. As uncontrolled buses inevitably bunch, active control is necessary to keep buses staggered. There are several methods described in the literature to avoid bunching and keep a staggered configuration \cite{Abkowitz1984,rossetti1998,Eberlein2001,Hickman2001,Fu2002,bin2006,Mukai2008,Daganzo2009,Cortes2010,cats2011,Gershenson2011,bartholdi2012,Chen2015,Ibarra-Rojas2015, Chen2016,Moreira-Matias2016,Wang2018,Alesiani2018,Menda2019, Saw2019intelligent,cats2011,Quek2020,Saw2019,Wang2020}. The final result in terms of the average waiting time depends on the specific dynamic control technique used, so the average waiting time determined here is an approximation. In section \ref{ssec:simulation} we compare the analytical calculations with a time-based simulation, employing a headway-based holding control. We see very similar results in terms of waiting times in all but the most extreme cases with very high $k$ where this approximation is less valid.
In Ref. \cite{Vismara2021conf} we show that bunching can happen within a single revolution if the value of $k$ is high enough that $n^* < 1$, where $n^*$ is the number of revolutions needed for buses to bunch, hence invalidating the assumption that buses can stay staggered.
Two buses starting from a perfectly staggered position in a loop with a single \emph{regular bus stop} bunch in $n^* = \log (k/2)/(2 \log (1-k))$ revolutions, according to Ref. \cite{Vismara2021conf}. For $k \ge 0.5$, uncontrolled buses bunch within the first revolution as $n^* \le 1$. In this regime, keeping the buses staggered is very challenging, if not impossible. In Fig. \ref{fig:sim}, the mismatch between a numerical simulation with dynamic control and the analytical results for the average waiting time is very significant when $k T$ approaches $0.5 T$, as buses cannot avoid bunching in the simulation.

The waiting time at the \emph{regular bus stop} is computed as half the longest waiting time, in the same way as it was in sections \ref{ssec:A} and \ref{ssec:B}, but here the buses are assumed to be staggered, hence the bus stop is served $N$ times every revolution. 
\begin{equation}\label{eq:w_s_C}
    \text{W}^\square_C = \frac{1}{2} \left( \frac{\bar{T}_C}{N} - \tau^\square_C \right) = \frac{1}{2} \frac{\bar{T}_C}{N} \left( 1 - k \right). 
\end{equation}
While boarding passengers from a bus stop, if other buses move, the system shifts away from a staggered configuration. To compute the average waiting time, we need to decide when the buses are considered perfectly staggered: either before or after they serve a bus stop. If buses are considered perfectly staggered \emph{before} serving the bus stop, the expression for $\text{W}^\square_C$ is the one in Eq. \eqref{eq:w_s_C}. However, if buses are considered staggered \emph{after} serving the bus stop, the average waiting time is $\tau^\square_C / 2$ longer than Eq. \eqref{eq:w_s_C}. We choose the first option, as it leads to lower waiting time. Different techniques of dynamic control to keep the buses staggered could be better described by either of the two choices of buses staggered \emph{before} or \emph{after} serving a bus stop. This difference is significant in the regime of high $\tau^\square_C = k \bar{T}_C / N$ where the assumption that buses can be staggered also fails.
Each revolution will take $\bar{T}_C$ as defined in Eq. \eqref{eq:tbar_C}.
The number of passengers boarding from the \emph{regular bus stop} during a revolution is computed as
\begin{equation}\label{eq:p_s_C}
    \text{Ppl}^\square_C = k \bar{T}_C,
\end{equation}
similarly to cases A and B in Eqs. \eqref{eq:p_s_A} \eqref{eq:p_s_B}.

The average waiting time at the \emph{spike bus stop} is calculated under the assumption that buses are fully staggered and uniformly distributed around the bus loop, hence the closest bus to the bus stop, when the spike arrives, is at an average distance (in units of time) $\delta^\triangle_{\text{closest}} = \bar{T}_C/(2 N)$ since each bus is separated by $\bar{T}_C/N$ unit times.
\begin{equation}\label{eq:w_t_C}
    \text{W}^\triangle_C = \delta^\triangle_{\text{closest}} + \frac{P}{2} = \frac{\bar{T}_C}{2 N} + \frac{P}{2}.
\end{equation}
The last term $P/2$ accounts for the time it takes to board the $P$ passengers, since the last one to board has to wait an additional $P$ units of time.
The quantity $\text{W}^\triangle_C$ has to be weighted by the average number of passengers boarding from the \emph{spike bus stop}:
\begin{equation}\label{eq:p_t_C}
    \text{Ppl}^\triangle_C = P \frac{\bar{T}_C}{T_s}.
\end{equation}

Finally, the average total time taken to complete a loop $\bar{T}_C$ is the sum of $T$ and the average dwelling time at the bus stops:
\begin{equation}\label{eq:tbar_C}
     \bar{T}_C = T + \tau^\triangle_C + \tau^\square_C = T + \frac{P \bar{T}_C}{T_s} + \frac{k \bar{T}_C}{N}
    = \frac{T}{1 - \frac{P}{T_s} - \frac{k}{N}}.
\end{equation}
The dwelling time at the \emph{spike bus stop} $\tau^\triangle_C$ is counted even though only one of the $N$ buses boards passengers there. The reason for including $\tau^\triangle_C$ is to account for active corrective actions by other buses, such as holding or slowing down, to maintain the perfectly staggered state. If that is not compensated for, the bus boarding at the \emph{spike bus stop} would have a slower revolution as compared with the other buses, causing buses to not be staggered anymore.
The average waiting time for passengers in the bus loop is computed as the weighted average of the waiting time at the two bus stops, Eqs. \eqref{eq:w_s_C} \eqref{eq:w_t_C}.
\begin{equation}\label{eq:w_C}
    \text{W}_C = \frac{\text{Ppl}^\triangle_C \times \text{W}^\triangle_C + \text{Ppl}^\square_C \times \text{W}^\square_C}{\text{Ppl}^\triangle_C + \text{Ppl}^\square_C} = \frac{N P^2 + \bar{T}_C \left(P  + k \left( 1 - k \right) T_s \right)}{2 N \left( P + k T_s\right)}.
\end{equation}


\subsection{Generalisation for more bus stops}\label{ssec:generalisation}
In this section \ref{sec:3unlimited} so far we have considered a scenario with only two bus stops for simplicity. It is possible to study a system with multiple \emph{regular bus stops} with our technique.  The general formula for the average waiting time with $M$ \emph{regular bus stops} and one \emph{spike bus stop} is
\begin{equation}\label{eq:w_general}
    \text{W} = \frac{\text{Ppl}^\triangle \times \text{W}^\triangle +\sum_{i=1}^{M} \text{Ppl}^{\square_i} \times \text{W}^{\square_i}}{\text{Ppl}^\triangle + \sum_{i=1}^{M} \text{Ppl}^{\square_i}} .
\end{equation}
The quantities to compute are the average waiting time at bus stop $i$ for all the $M$ \emph{regular bus stops} $\text{W}^{\square_i}$ and the people boarding from there $\text{Ppl}^{\square_i}$. Moreover, stopping in multiple bus stops means that the total time taken to complete a loop, $\bar{T}$, will need to take into account the extra time taken to board from more than one \emph{regular bus stops}, except for the case of synchronised bunched buses in section \ref{ssec:B} where the total time to complete a loop is the period of the spike $T_s$.
In the following part, we present the general expressions for the waiting time at the $M$ \emph{regular bus stops}, each of which with demand $s_i = k_i \times l$ for the three cases examined: bunched buses, synchronised bunching, perfectly staggered buses.

\subsubsection{Bunched buses}
Using the same reasoning as in Eq. \eqref{eq:w_s_A}, the waiting time at each \emph{regular bus stops} in the case of $N$ bunched buses is:
\begin{equation}\label{eq:w_s_A_i}
	\text{W}^{\square_i}_A = \frac{1}{2} \left( \bar{T}_A - \tau^{\square_i}_A \right) = \frac{\bar{T}_A}{2} \left(1 - \frac{k_i}{N}\right).
\end{equation}
The passengers boarding at each bus stops are:
\begin{equation}\label{eq:p_s_A_i}
	\text{Ppl}^{\square_i}_A = k_i \bar{T}_A.
\end{equation}
The total time taken to complete the loop is:
\begin{equation}\label{eq:tbar_A_i}
	\bar{T}_A = T + \tau^\triangle_A + \sum_{i=1}^{M} \tau^{\square_i}_A = T + \frac{P \bar{T}_A}{N T_s} + \sum_{i=1}^{M} \frac{k_i \bar{T}_A}{N} .
\end{equation}
The waiting time at the \emph{spike bus stop} is indirectly affected by $\bar{T}_A$.
The average waiting time of passengers in a loop with $M$ \emph{regular bus stops} and one \emph{spike bus stop} is computed via Eq. \eqref{eq:w_general} by combining Eqs. \eqref{eq:w_s_A_i}, \eqref{eq:p_s_A_i} and \eqref{eq:tbar_A_i} along with $\text{W}^{\triangle}_A$ and $\text{Ppl}^{\triangle}_A$ from Eqs. \eqref{eq:w_t_A} \eqref{eq:p_t_A}.

\subsubsection{Synchronised bunched buses}
Following the idea used in section \ref{ssec:B}, the waiting time at each \emph{regular bus stops} in the case of $N$ synchronised bunched buses is:
\begin{equation}
	\text{W}^{\square_i}_B = \frac{T_s - \tau^{\square_i}_B}{2} = \frac{T_s}{2} \left(1 - \frac{k_i}{N}\right).
\end{equation}
The passengers boarding at each bus stops are:
\begin{equation}
	\text{Ppl}^{\square_i}_B = k_i T_s.
\end{equation}
The total time taken to complete the loop is still $T_s$, provided that the extra dwelling time does not slow down the revolution of the buses below the period of the spike $T_s$.
The waiting time at the \emph{spike bus stop} is not affected by the extra bus stops.
By combining the equations in this section with Eqs. \eqref{eq:w_t_B} \eqref{eq:p_t_B} and substituting them in Eq. \eqref{eq:w_general}, we can compute the average waiting time of passengers in a loop with $M$ \emph{regular bus stops} and one \emph{spike bus stop} in the case of synchronised bunched buses, generalising the result in section \ref{ssec:B}.

\subsubsection{Perfectly staggered buses}
Analogously as how it is calculated in Eq. \eqref{eq:w_s_C}, the waiting time at a \emph{regular bus stops} in the case of $N$ perfectly staggered buses is:
\begin{equation}
	\text{W}^{\square_i}_C = \frac{1}{2} \frac{\bar{T}_C}{N} \left( 1 - k \right).
\end{equation}
The passengers boarding at each bus stops are:
\begin{equation}
	\text{Ppl}^{\square_i}_C = k_i \bar{T}_C.
\end{equation}
The total time taken to complete the loop is:
\begin{equation}
	\bar{T}_C = T + \tau^\triangle_C + \sum_{i=1}^{M} \tau^{\square_i}_C = T + \frac{P \bar{T}_C}{T_s} + \sum_{i=1}^{M} \frac{k_i \bar{T}_C}{N} .
\end{equation}
The waiting time at the \emph{spike bus stop} is indirectly affected by $\bar{T}_C$ but takes the same functional form as in Eq. \eqref{eq:w_t_C}. Combining the equations in this section along with Eq. \eqref{eq:p_t_C} and substituting them in Eq. \eqref{eq:w_general} it is possible to compute the average waiting time of passengers in a loop with $M$ \emph{regular bus stops} and one \emph{spike bus stop} in the case of perfectly staggered buses, generalising the result in section \ref{ssec:C}.


\section{The effect of limited capacity for the buses}\label{sec:limited}
Real-world transportation systems have limited capacity regarding the number of commuters that can board. Such limitations can alter the optimal dispatch of vehicles and headway \cite{Newell1971, Sadrani2021, Niu2013, Zhu2017}, hence staggered solutions are not necessarily optimal even in the presence of only \emph{regular bus stops}. Ref. \cite{Shi2020} deals more explicitly with bunching buses, in the form of a newly proposed modular vehicle system. The modular vehicles can be combined to increase capacity, which is similar to the effect of bunched buses. The system however is studied in the context of a splitting route where different modules travel to different destinations: a passenger has to board a specific module and none of the cases cited above deals with loops or \emph{spike bus stops}.

In this section, we only consider the capacity limit at the \emph{spike bus stop}. All of the buses can board up to $Q$ passengers at the \emph{spike bus stop}. For dimensional consistency with our convention for $P = p / l$, we consider capacity over the fixed boarding rate $l$ to simplify our equation with the quantity $c = Q / l$.
\subsection{Bunched buses}\label{ssec:A_lim}
One of the advantages of bunched buses is the cumulative capacity of the platoon, which grows linearly with the number of buses. Limited capacity affects this scenario only when $c < P / N$, if $c \ge P / N$ the result in Eq. \eqref{eq:w_A} applies. In this section we present the result in the regime $P / (2 N) < c < P / N$. In this case, the buses cannot board all of the passengers at the \emph{spike bus stop}, so they need to complete another revolution to board passengers from there. For that to be possible, the spikes cannot be too frequent, so a necessary condition is $T_s \ge 2 \bar{T}_{A_\text{lim}}$, 
where $\bar{T}_{A_\text{lim}}$ is the total time needed to complete a revolution.
If the inequality above is not satisfied, more passengers are arriving than those that the buses can pick up, hence the number of passengers waiting will grow at every revolution, leading to a diverging waiting time.

The major difference, as compared with the case with unlimited capacity in section \ref{ssec:A}, is the waiting time at the \emph{spike bus stop}.
\begin{equation}\label{eq:w_t_A_lim}
\begin{split}
	\text{W}^\triangle_{A_\text{lim}} &= \frac{N c}{P} \left( \delta^\triangle_\text{first} + \frac{N c}{2 N} \right) + \frac{P - N c}{P} \left( \delta^\triangle_\text{second} + \frac{P - N c}{2 N} \right) = \\
    &= \frac{N c}{P} \left( \frac{\bar{T}_{A_\text{lim}}}{2} + \frac{N c}{2 N} \right) + \frac{P - N c}{P} \left( \frac{3 \bar{T}_{A_\text{lim}}}{2} + \frac{P - N c}{2 N} \right).
\end{split}
\end{equation}
Following the same idea as in section \ref{ssec:A}, we define $\delta^\triangle_\text{first}$ and $\delta^\triangle_\text{second}$ as the distance (in units of time) of the buses when they serve the \emph{spike bus stop} for the first and second time, respectively. Being the buses bunched, the second time the buses pick up passengers in the \emph{spike bus stop} is a whole revolution after the first time, therefore there is a difference of $\bar{T}_{A_\text{lim}}$ between $\delta^\triangle_\text{first}$ and $\delta^\triangle_\text{second}$. The waiting time of passengers picked up by the buses between the two revolutions is weighted by the passengers boarded, $N c$ at the first revolution and $P - N c$ in the second round.
The average number of passengers boarded per revolution is
\begin{equation}\label{eq:p_t_A_lim}
	\text{Ppl}^\triangle_{A_\text{lim}} = P \frac{\bar{T}_{A_\text{lim}}}{T_s},
\end{equation}
as $P$ passengers arrive once every $T_s / \bar{T}_{A_\text{lim}}$ revolutions.

The values of $\text{W}^\square_{A_\text{lim}}$ and $\text{Ppl}^\square_{A_\text{lim}}$ are the same as in Eqs. \eqref{eq:w_s_A} \eqref{eq:p_s_A}, with $\bar{T}_{A}$ replaced by $\bar{T}_{A_\text{lim}}$: $\text{W}^\square_{A_\text{lim}} = \bar{T}_{A_\text{lim}}/2 (1 - k / N )$ and $\text{Ppl}^\square_{A_\text{lim}} = k \bar{T}_{A_\text{lim}} $.
To compute the average time taken to complete a revolution, $\bar{T}_{A_\text{lim}}$, we consider the average time spent at the \emph{spike bus stop} during a revolution. Every $T_s / \bar{T}_{A_\text{lim}}$ revolutions, a spike happens, and for every spike the buses need to stop to pick up $N c$ and $P - N c$ passengers at the first and second revolution respectively. Hence the average time to complete a revolution is
\begin{equation}\label{eq:tbar_A_lim}
\begin{split}
	\bar{T}_{A_\text{lim}} &= T + \tau^{\triangle (1)}_{A_\text{lim}} + \tau^{\triangle (2)}_{A_\text{lim}} + \tau^\square_{A_\text{lim}}
    = T + \frac{\bar{T}_{A_\text{lim}}}{T_s} \frac{N c}{N} + \frac{\bar{T}_{A_\text{lim}}}{T_s} \frac{P - N c}{N} + \frac{k \bar{T}_{A_\text{lim}}}{N} = \\
    &= \frac{T}{1 - \frac{P}{N T_s} - \frac{k}{N}}.
\end{split}
\end{equation}

Combining Eqs. \eqref{eq:w_t_A_lim} \eqref{eq:p_t_A_lim} \eqref{eq:w_s_A} \eqref{eq:p_s_A} \eqref{eq:tbar_A_lim}, the average waiting time for passengers in a scenario with bunched buses and reduced capacity $P / (2 N) < c < P / N$ is:
\begin{equation}
    \text{W}_{A_\text{lim}} = \frac{N \bar{T}_{A_\text{lim}} \left( 3 P - 2 N c + k T_s \left(1 - \frac{k}{N}\right)\right) + \left(N c\right)^2 + \left(P - N c\right)^2}{2 N \left( P + k T_s \right)}.
\end{equation}

It is possible to generalise the results for $c < P / (2 N)$ by adding extra terms in Eq. \eqref{eq:w_t_A_lim} using $\delta^\triangle_\text{i-th} = \bar{T}_{A_\text{lim}}/2 \times (2 i - 1)$. The value of $\bar{T}_{A_\text{lim}}$ does not change since it does not depend on $c$, as seen in Eq. \eqref{eq:tbar_A_lim}.

\subsection{Synchronised bunched buses}\label{ssec:B_lim}
Similarly to the bunched configuration in section \ref{ssec:A_lim}, limiting the capacity of the buses in this synchronised bunched setting affects the dynamics and the waiting time of buses only in the regime $c < P / N$. In this section we consider only the case of $P / (2 N) < c < P / N$ where the platoon of bunched buses waits for the spike of passengers at the \emph{spike bus stop}, but the capacity is not enough to board them all, so another revolution is needed to pick up the remaining passengers.
We also assume that the buses can always board all of the passengers at the \emph{regular bus stop}.
In the same way as the case in section \ref{ssec:A_lim}, the spikes need to have a long enough period $T_s$ in such a way that the buses can pick up all the passengers at the spike bus stop before a new spike happens. The formal condition is $T_s \ge  \bar{T}^{(1)}_{B_\text{lim}} +  \bar{T}^{(2)}_{B_\text{lim}}$ where $ \bar{T}^{(1)}_{B_\text{lim}} $ represents the time taken to complete the first revolution after the spike and $ \bar{T}^{(2)}_{B_\text{lim}} $ is the time taken to complete the second revolution, just before holding $\text{Hold}_{B_\text{lim}}$ unit times at the \emph{spike bus stop}, waiting for the new spike. Those quantities are defined in Eqs. \eqref{eq:tbar1_B_lim} and \eqref{eq:tbar2_B_lim} and related to $T_s$ according to the equation $T_s = \bar{T}^{(1)}_{B_\text{lim}} + \bar{T}^{(2)}_{B_\text{lim}} + \text{Hold}_{B_\text{lim}}$.

To compute the waiting time, we can break the dynamics into three parts: first revolution, second revolution and holding to synchronise with the next spike. The first part starts from the arrival of passengers at the \emph{spike bus stop}, where the platoon of buses is waiting to board them. The first part ends when the \emph{spike bus stop} is reached a second time. In our equations it is indicated with the $(1)$ notation and it lasts $\bar{T}^{(1)}_{B_\text{lim}}$ unit times. The second part starts when the platoon of buses picks up the remaining passengers at the \emph{spike bus stop} and it ends a revolution later, when the buses reach the \emph{spike bus stop} again, just before starting to hold to wait for the next spike.We indicate the quantities relative to this period with the index $(2)$ and this revolution lasts $\bar{T}^{(2)}_{B_\text{lim}}$ unit times. The final phase consists of the buses holding at the \emph{spike bus stop} waiting for a new spike to occur.

To compute the waiting time at the \emph{regular bus stop}, we average the waiting time the first and the second time the buses reach this bus stop. At the first revolution $(1)$, the bus stop has not been served for a time $\text{Hold}_{B_\text{lim}} + \bar{T}^{(1)}_{B_\text{lim}} -  \tau^{\square (1)}_{B_\text{lim}} = T_s - \bar{T}^{(2)}_{B_\text{lim}} -  \tau^{\square (1)}_{B_\text{lim}}$ and the average waiting time, under the hypothesis of constant arrival at rate $s = k \times l$, is half the longest time waited by a passenger, hence
\begin{equation}
	\text{W}^{\square (1)}_{B_\text{lim}} = \frac{1}{2} \left( T_s - \bar{T}^{(2)}_{B_\text{lim}} - \tau^{\square (1)}_{B_\text{lim}} \right) = \frac{T_s-\bar{T}^{(2)}_{B_\text{lim}}}{2} \left( 1 - \frac{k}{N}\right).
\end{equation}
The passengers boarding in the first loop are $\text{Ppl}^{\square (1)}_{B_\text{lim}} = k (T_s-\bar{T}^{(2)}_{B_\text{lim}})$. The same procedure applies to the waiting time of the second period $(2)$ where the \emph{regular bus stop} has not been served for $\bar{T}^{(2)}_{B_\text{lim}} - \tau^{\square (2)}_{B_\text{lim}}$ units of time.
\begin{equation}
	\text{W}^{\square (2)}_{B_\text{lim}} = \frac{1}{2} \left(\bar{T}^{(2)}_{B_\text{lim}} - \tau^{\square (2)}_{B_\text{lim}} \right) = \frac{\bar{T}^{(2)}_{B_\text{lim}}}{2} \left( 1 - \frac{k}{N}\right),
\end{equation}
with $\text{Ppl}^{\square (2)}_{B_\text{lim}} = k \bar{T}^{(2)}_{B_\text{lim}}$ passengers boarded, for a total of $\text{Ppl}^{\square}_{B_\text{lim}}  = k T_s$ passengers boarding from the \emph{regular bus stop} during the two revolutions.

The waiting time at the \emph{spike bus stop} also needs to be calculated averaging the waiting time $\text{W}^{\triangle (1)}_{B_\text{lim}}$ of the first $\text{Ppl}^{\triangle (1)}_{B_\text{lim}} = N c$ passengers that can board immediately after the spike during revolution $(1)$, so only the time taken to board is considered in the waiting time, and the waiting time $\text{W}^{\triangle (2)}_{B_\text{lim}}$ of the remaining $\text{Ppl}^{\triangle (2)}_{B_\text{lim}} = P - N c$ that needs to wait $\bar{T}^{(1)}_{B_\text{lim}}$ until the first revolution is completed. The average waiting time at the \emph{spike bus stop} is computed as
\begin{equation}
\begin{split}
	\text{W}^{\triangle (1)}_{B_\text{lim}} &=  \frac{N c}{2 N},\\
	\text{W}^{\triangle (2)}_{B_\text{lim}} &= \bar{T}^{(1)}_{B_\text{lim}} + \frac{P- N c}{2 N},\\
	\text{W}^\triangle_{B_\text{lim}} &= \frac{N c}{P} \text{W}^{\triangle (1)}_{B_\text{lim}} + \frac{P- N c}{P} \text{W}^{\triangle (2)}_{B_\text{lim}}.
\end{split}
\end{equation}

The waiting time computed requires an expression for the time taken to complete the revolutions $(1)$ and $(2)$. The first case is expressed as:
\begin{equation}\label{eq:tbar1_B_lim}
	\bar{T}^{(1)}_{B_\text{lim}} = T + \tau^{\triangle (1)}_{B_\text{lim}} + \tau^{\square (1)}_{B_\text{lim}} = T + c + \frac{k}{N} \left(T_s - \bar{T}^{(2)}_{B_\text{lim}} \right).
\end{equation}
The revolution includes $T$ unit times on the road, $\tau^{\triangle (1)}_{B_\text{lim}}$ to board form the spike bus stop the first $N c$ passengers, and $\tau^{\square (1)}_{B_\text{lim}}$ to board passengers at the \emph{regular bus stop}. Passengers at the \emph{regular bus stop} started arriving during the previous $(2)$ revolution, after the platoon of buses finished to board at the \emph{regular bus stop}.
The number of passengers arriving at the \emph{regular bus stop} during the first revolution $(1)$ (and the dwelling time $\tau^{\square (1)}_{B_\text{lim}}$) is proportional to the sum of the holding time waited to synchronise with the spike after revolution $(2)$ and $\bar{T}^{(1)}_{B_\text{lim}}$. The quantity $\bar{T}^{(1)}_{B_\text{lim}}$ is defined as the time taken to board passengers at the \emph{spike bus stop}, the time taken to board up to the last passengers at the \emph{regular bus stop} during the first revolution $(1)$, and the time to complete the loop itself, $T$: $\tau^{\square (1)}_{B_\text{lim}} \propto \text{Hold}_{B_\text{lim}} + \bar{T}^{(1)}_{B_\text{lim}}$. As the duration of the two revolutions $(1)$ and $(2)$ plus the time $\text{Hold}_{B_\text{lim}}$ waited at the \emph{spike bus stop} must be $T_s$ in order to synchronise with the spike, $\text{Hold}_{B_\text{lim}} + \bar{T}^{(1)}_{B_\text{lim}} = T_s - \bar{T}^{(2)}_{B_\text{lim}}$.

The time taken to complete the second revolution $(2)$ before holding is:
\begin{equation}\label{eq:tbar2_B_lim}
	\bar{T}^{(2)}_{B_\text{lim}} = T + \tau^{\triangle (2)}_{B_\text{lim}} + \tau^{\square (2)}_{B_\text{lim}} = T + \frac{P - N c}{N} + \frac{k}{N} \bar{T}^{(2)}_{B_\text{lim}} = \frac{T +  \frac{P - N c}{N}}{1-\frac{k}{N}}.
\end{equation}
The dwelling time taken to board at the \emph{spike bus stop} is proportional to the number of passengers left $P - N c$. For the case of the \emph{regular bus stop}, the number of passengers, hence the dwelling time, are proportional to $ \bar{T}^{(2)}_{B_\text{lim}} $ itself, as there is no waiting for synchronisation in between serving the \emph{regular bus stop} during the first $(1)$ and second $(2)$ revolution in this setting.

Combining the above expressions with $\text{Ppl}^{\square}_{B_\text{lim}}  = k T_s$ and $\text{Ppl}^{\triangle}_{B_\text{lim}}  = P$, the average waiting time for synchronised bunched buses with capacity $P / (2 N) < c < P / N$ is found from
\begin{equation}
	\text{W}_{B_\text{lim}}  = \frac{\text{Ppl}^\triangle_{B_\text{lim}}  \times \text{W}^\triangle_{B_\text{lim}}  + \text{Ppl}^\square_{B_\text{lim}}  \times \text{W}^\square_{B_\text{lim}} }{\text{Ppl}^\triangle_{B_\text{lim}}  + \text{Ppl}^\square_{B_\text{lim}}}.
\end{equation}

The average waiting time for synchronised bunched buses is equal to that of a single bus with capacity $N c$ and boarding rate $N l$. Assuming that it is possible to build and deploy such a bus, $N$ synchronised bunched buses are still more versatile and adaptable to a change in demand. Firstly, $N$ buses can implement active control and move to a staggered configuration when the passenger demand at the bus stops changes. As we see from Figs. \ref{fig:k_comparison} and \ref{fig:p_c}, the more advantageous configuration can be either perfectly staggered buses or synchronised bunched buses, depending on the parameters of the bus loop. Another advantage of $N$ bunched buses over a single bus with equivalent capacity and boarding rate is the option of adding and removing extra buses to match the passenger demand and capacity constraints as the number of commuters in the system changes, reducing operational cost when less capacity is needed.

\subsection{Perfectly staggered buses}\label{ssec:C_lim}
As the \emph{spike bus stop} is being served by one bus at a time in this configuration of perfectly staggered buses, the limited capacity affects the system for any $c < P$. In the previous two cases in sections \ref{ssec:A_lim} and \ref{ssec:B_lim}, the limited capacity affects the system only for $c < P / N$. 

We consider the case where $m$ buses are needed to board the total number of passengers at the \emph{spike bus stop} ($P/m<c<P/(m-1)$). The value for $m$ is $\lceil P / c \rceil $.
As this case requires $m/N$ revolutions to clear the passengers at the spike bus stop, a necessary condition for the buses to serve this system is $T_s > (m / N) T$
, otherwise the number of waiting commuters blows up.
Following the same reasoning as in Eqs. \eqref{eq:w_t_A_lim} and \eqref{eq:w_t_C},
\begin{equation}\label{eq:w_t_C_lim}
\begin{split}
    \text{W}^\triangle_{C_\text{lim}} &= \frac{c}{P} \left( \delta^\triangle_{\text{1st}} + \frac{c}{2}\right) +
    \frac{c}{P} \left( \delta^\triangle_{\text{2nd}} + \frac{c}{2}\right) + \\
    & + \dots + \frac{P - (m-1)c}{P}\left( \delta^\triangle_{\text{m-th}} + \frac{P-(m-1)c}{2} \right) ,
\end{split}
\end{equation}
where $ \delta^\triangle_{\text{i-th}} = \bar{T}_{C_\text{lim}} (2 i -1) / (2 N)$ is the distance (in units of time) of the i-th bus to serve the \emph{spike bus stop}.
As in the case with unlimited capacity, the number of people boarding is $\text{Ppl}^\triangle_{C_\text{lim}} = P \bar{T}_{C_\text{lim}} / T_s$.

The waiting time and people boarded from the \emph{regular bus stop} are also the same as in Eqs. \eqref{eq:w_s_C} and \eqref{eq:p_s_C}: $\text{W}^\square_{C_\text{lim}} = \bar{T}_{C_\text{lim}} / (2 N) (1 - k)$ 
and $\text{Ppl}^\square_{C_\text{lim}} = k \bar{T}_{C_\text{lim}} $.
To express the average waiting time in the whole system, the last remaining equation is for $\bar{T}_{C_\text{lim}}$. As in the case of unlimited capacity in Eq. \eqref{eq:tbar_C}, boarding passengers at the \emph{regular bus stop} takes $\tau^\square_{C_\text{lim}} = k \bar{T}_{C_\text{lim}} / N$ for each bus, under the assumption of perfectly staggered buses and no capacity constraints at the \emph{regular bus stop}. At the \emph{spike bus stop} buses will dwell for either $c$ or $P - (m-1) c$ unit times if there are passengers there. For perfectly staggered buses, the necessary condition is for all the buses to move with the same period, hence they all have to employ dynamic control to keep themselves staggered, slowing down at the pace of the slowest buses. Since $c > P - (m-1) c$, by definition of $m = \lceil P / c \rceil$, the slowest bus waits $c$ unit times, and that happens $T_s / \bar{T}_{C_\text{lim}}$ times a loop. The expression for the average time taken to complete a loop is then:
\begin{equation}\label{eq:tbar_C_lim}
	\bar{T}_{C_\text{lim}} = T + \tau^\triangle_{C_\text{lim}} + \tau^\square_{C_\text{lim}} = T + \frac{c \bar{T}_{C_\text{lim}}}{T_s} + \frac{k \bar{T}_{C_\text{lim}}}{N}
    = \frac{T}{1 - \frac{c}{T_s} - \frac{k}{N}}.
\end{equation}

The general expression depends on the value of $m$ via $\text{W}^\triangle_{C_\text{lim}}$. We report the average waiting time in the case of $m = 2$, hence in the regime $P / 2 < c < P$, but the equations above allow for a general calculation via Eq. \eqref{eq:w_general}.
\begin{equation}
    \text{W}_{C_\text{lim}} = \frac{N \left( c^2 + \left( P - c \right)^2 \right) + \bar{T}_{C_\text{lim}} \left( 3 P - 2 c + k \left( 1 - k \right) T_s \right)}{2 N \left( P + k T_s \right)}.
\end{equation}

\section{Comparison and discussion}\label{sec:comparison}
In a bus loop with a \emph{spike bus stop} and a \emph{regular bus stop} six important variables affect the system: the crowdedness of the bus stops, through $P$ and $k$, the period of the spikes $T_s$, the period for a revolution without dwelling $T$, the number of buses $N$ and the capacity of buses $c$.

Firstly, as the perfectly staggered buses are the best solution in terms of waiting time for a \emph{regular bus stop} \cite{Vismara2021conf, Azfar2018}, we expect that for lower demand of the \emph{spike bus stop} as compared with the \emph{regular bus stop}, the staggered configuration will outperform the two bunched configurations examined in this work. Computing the average waiting time of the three cases with unrestricted capacity in the limit $P = 0$ in Eqs. \eqref{eq:w_A}, \eqref{eq:w_B} and \eqref{eq:w_C} we have:
\begin{align}
    \text{W}_A(P=0) &= \bar{T}_A \left(1 - \frac{k}{N} \right) = \frac{T}{2}, \\ 
    \text{W}_B(P=0) &= \frac{T_s}{2} \left(1 - \frac{k}{N} \right), \\
    \text{W}_C(P=0) &= \frac{\bar{T}_C}{2 N} \left(1 - k \right) = \frac{T}{2} \frac{1 - k}{N - k} .
\end{align}
The waiting time for bunched buses, case $A$, is fixed as $T/2$ since as long as passengers can be boarded faster than they arrive ($k < N$), no passenger has to wait for more than a whole revolution $T$ to board.
The synchronised bunched buses in case $B$ behave in the same way, except that having to wait at the \emph{spike bus stop}, 
the time taken to complete a revolution is $T_s \ge \bar{T}_A$, where the inequality is a necessary condition for synchronised bunched buses, so $\text{W}_A(P=0) \le \text{W}_B(P=0)$.
In the last case of perfectly staggered buses, $C$, the waiting time is always the lowest, as expected. The quantity $(1-k)/(N-k)$ is always less or equal than $1$, since $k < 1$ is a necessary condition for staggered buses, so $\text{W}_C(P=0) \le \text{W}_A(P=0) \le \text{W}_B(P=0)$. In the case of a single bus $N=1$, case $A$ and case $C$ are equivalent.
Taking the limit $P=0$ leads to the same result as taking the limit for $T_s \to \infty$ for the case of bunched buses $A$ and perfectly staggered buses $C$. For synchronised bunched buses, this limit does not make sense as the buses would have to wait indefinitely at the \emph{spike bus stop}.

In the opposite limit where all demand is concentrated at the \emph{spike bus stop} ($k=0$) we expect the synchronised bunched buses to outperform the other configurations as the platoon of buses waits for the spike to arrive, eliminating any delay between arrival and beginning of boarding process. From Eqs. \eqref{eq:w_A}, \eqref{eq:w_B} and \eqref{eq:w_C}, in the setting of unconstrained capacity:
\begin{align}
    \text{W}_A(k=0) &= \frac{P + N \bar{T}_A}{2 N} = \frac{P}{2 N} + \frac{T}{2} \frac{1}{1 - \frac{P}{N T_s}}, \label{eq:w_A_k0}\\
    \text{W}_B(k=0) &= \frac{P}{2 N}, \\
    \text{W}_C(k=0) &= \frac{N P + \bar{T}_C}{2 N} = \frac{P}{2} + \frac{T}{2 N} \frac{1}{1 - \frac{P}{T_s}} \label{eq:w_C_k0}.
\end{align}
It takes $P$ unit times to board $P$ passengers for perfectly staggered buses $C$, so if another spike occurs at $T_s \le P$, the number of passengers waiting grows with time, either diverging or causing the buses to bunch, against the hypothesis of perfectly staggered buses, hence we consider $T_s \ge P$. Synchronised buses $B$ always perform the best among the three, since $1 - P/(N T_s)$ and $1 - P / T_s$ are necessarily positive. For cases $A$ and $C$, in the limit of small $P \ll T < T_s$, the waiting time is dominated by the time needed for a bus to reach the bus stop, so bunched buses have the longest waiting time as, on average, buses are further away from the \emph{spike bus stop} $\text{W}_A(k=0, P \ll T) \approx T/2 > \text{W}_C(k=0, P \ll T) \approx T/(2 N)$. In the opposite limit, where $P \to T_s$, perfectly staggered buses cannot keep up with the demand while staying staggered, hence $\text{W}_C(k=0, P \to T_s) \to \infty$. A similar analysis can be performed in the case of constrained capacity $c$ from the equations of the waiting time in section \ref{sec:limited}.

\begin{figure}[ht]
    \centering
    \includegraphics[width=15cm]{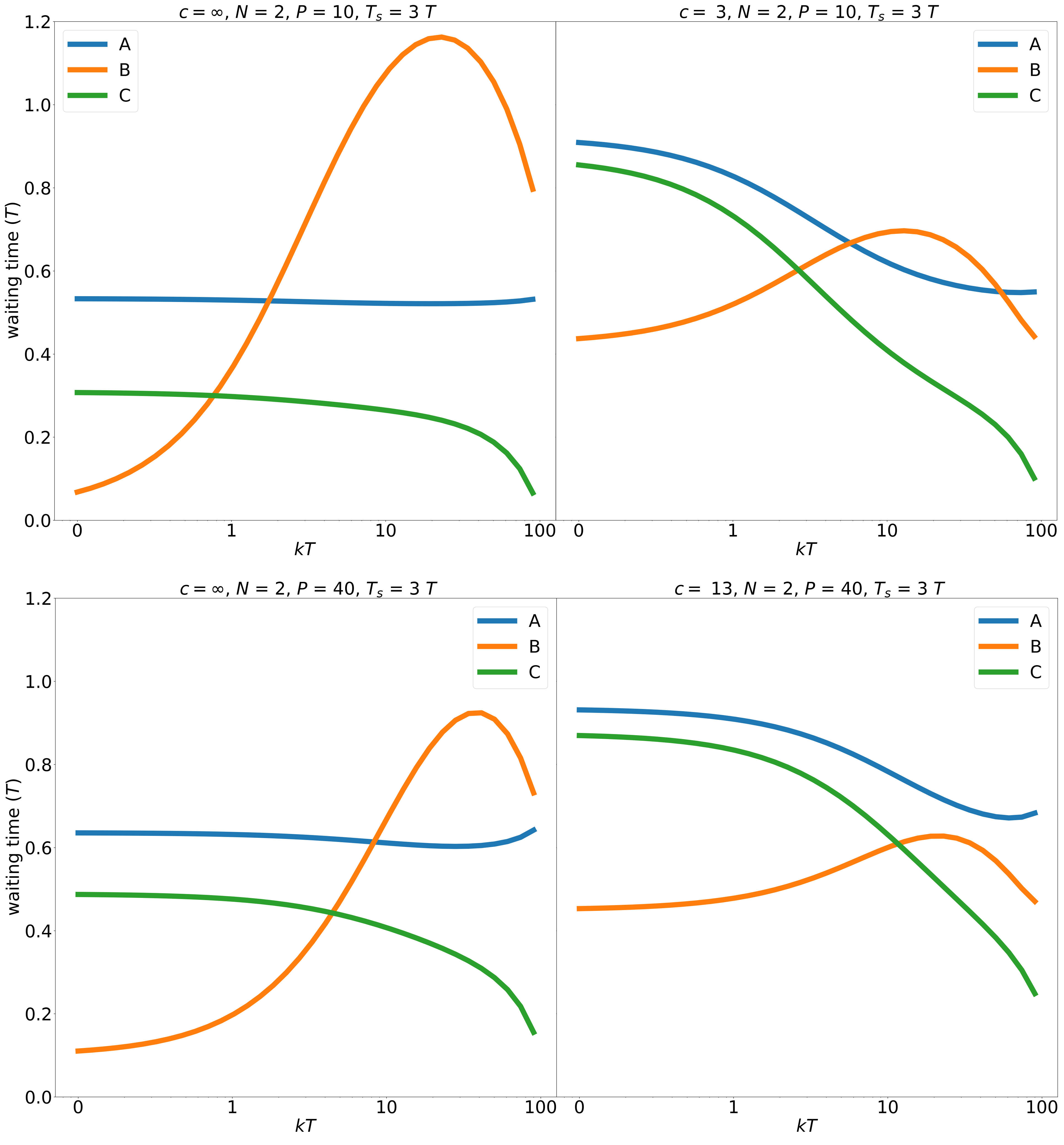}
    \caption{Average waiting time for two buses in the three configurations $A, B, C$ described in sections \ref{sec:3unlimited} and \ref{sec:limited}. The two upper plots represent a case of low demand $P$ from the \emph{spike bus stop} while the two lower plots represent high demand $P$. The plots on the left assume unlimited capacity of the buses, the plots on the right explicitly limit the capacity of buses at $c = \lfloor P / 3 \rfloor$ while boarding passengers from the \emph{spike bus stop}. As discussed in section \ref{sec:comparison}, synchronised bunched buses (configuration $B$), outperform perfectly staggered buses (configuration $C$) when the demand of the \emph{spike bus stop} is high as compared with the \emph{regular bus stop}. Limiting capacity further increases the advantage of bunched buses.
    For this example, $T$ is set to a value of $100$ unit times and $T_s$ is three times $T$.}
    \label{fig:k_comparison}
\end{figure}

Intermediate cases of $k$ and $P$ are less intuitive, and introducing the capacity limit $c$ makes the expressions less simple. To understand and compare the behaviour of the buses in such a scenario, we visualise the effect that the aforementioned variables have on the system.
Firstly, the average waiting time as a function of $k$ is explored in Fig. \ref{fig:k_comparison} where the three methods (bunched buses, synchronised bunched buses and perfectly staggered buses) are compared in a case with high and low demand from the \emph{spike bus stop} $P$, in two conditions of unlimited and limited capacity. As expected from the analytical comparison above, synchronised bunched buses outperform perfectly staggered buses in the regime of low demand $k T$ from the \emph{regular bus stop}. The comparison is done with $k T$ to have dimensional consistency with $P$.
The curve for the synchronised bunched buses $B$ from Fig. \ref{fig:k_comparison} is particularly interesting. The average waiting time increases with $k$ for low values of $k$ as expected, but after reaching a peak, the waiting time decreases in the regime of high $k$. The downward trend can be explained from the equations \eqref{eq:w_s_B} and \eqref{eq:w_t_B}. The contribution to the average waiting time from the \emph{spike bus stop}, in the case of synchronised bunched buses $B$, does not depend on $k$, nor the number of passengers boarded from the \emph{spike bus stop} $P$. The number of passengers boarded at the \emph{regular bus stop} $\text{Ppl}^\square_B \propto k$ increases with $k$, while $P$ is constant, hence eventually the contribution at the \emph{regular bus stop} dominates for high $k$, showing the downward trend $\propto (1 - k/N)$ since $\text{W}^\square_B \propto (1 - k/N)$. An intuitive explanation is that at larger $k$ the buses spend more time at the \emph{regular bus stop} and less time waiting for the spike at the \emph{spike bus stop}.
In all of the cases examined in Fig. \ref{fig:k_comparison}, bunching buses $A$ are never the best performing configuration regarding minimising the average waiting time.

\begin{figure}[ht]
    \centering
        \includegraphics[width=16cm]{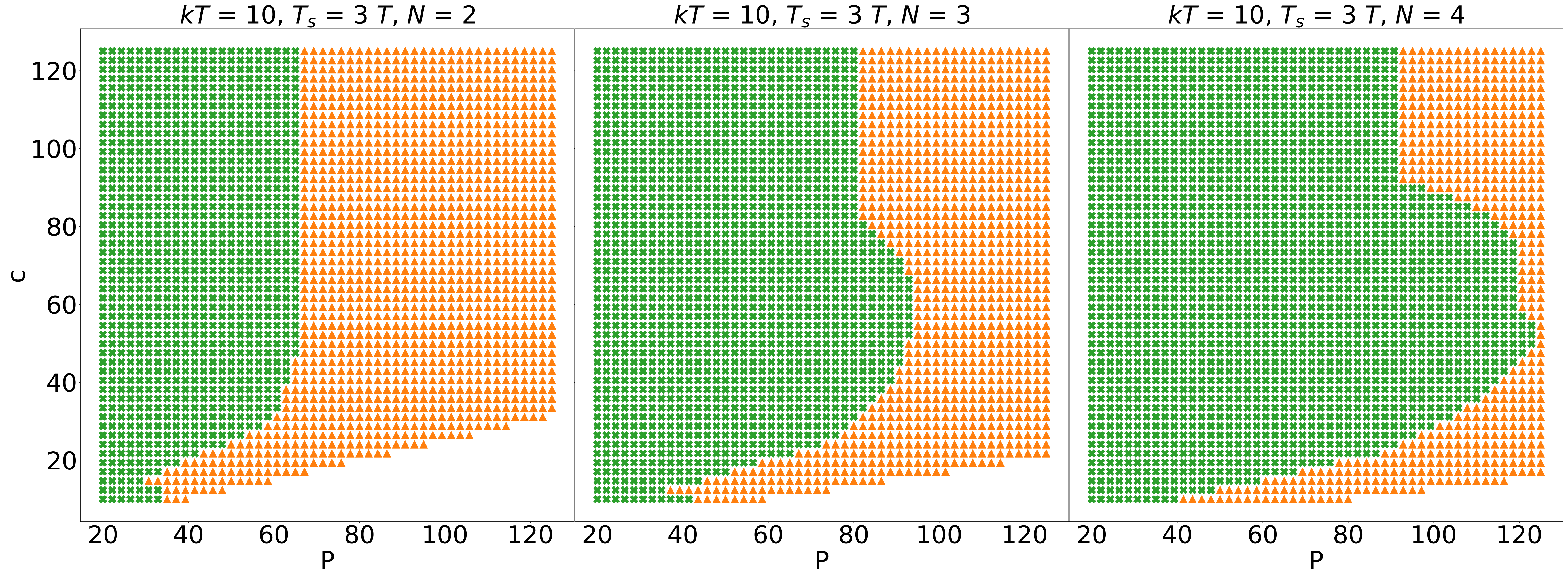}
    \caption{Comparison of the average waiting time for the three configurations as a function of the demand at the \emph{spike bus stop} $P$ and the capacity $c$. The orange triangles represent the set of parameters where synchronised bunched buses have the lowest average waiting time, while the green ''x`` symbols indicate where perfectly staggered buses have the lowest average waiting time among the three configurations. Bunched buses (case $A$) are never the best performers for the parameters explored. The three plots show how increasing the number of buses from 2 to 3 to 4 (from left to right) tends to favour the staggered bus configuration. Points in the region of high $P$ and low $c$ are missing since in that regime more than two revolutions are needed to pick up all the passengers at the \emph{spike bus stop}. This region is defined by $c < P / (2 N)$. For this example, $T$ is set to a value of $100$ unit times and $T_s$ is three times $T$ while $k$ is fixed at $0.1$.}
    \label{fig:p_c}
\end{figure}

The impact of the limited capacity is visualised in Fig. \ref{fig:p_c} where the three configurations of buses are compared, changing the number of buses $N$ from 2 to 4. In the range of parameters explored, bunched buses are always outperformed by synchronised bunched buses or perfectly staggered buses, in terms of average waiting time. While the case of 2 buses in the leftmost plot in Fig. \ref{fig:p_c} seems to follow the intuition that low capacity $c$ and high spike demand $P$ favour synchronised bunched buses over perfectly staggered buses, we see that for $N=3$ and $N=4$ the situation is less intuitive for intermediate values of $c$.
The explanation of this phenomenon has to do with the fact that lowering the capacity $c$ in the case of perfectly staggered buses, lowers the average time that buses take to complete a loop $\bar{T}_C$ since each bus has to stop less time at the \emph{spike bus stop}, according to Eq. \eqref{eq:tbar_C_lim}. This reduction in $\bar{T}_C$ is directly proportional to the reduction in waiting time at the \emph{regular bus stop} via Eq. \eqref{eq:w_s_C} and indirectly reduces the waiting time at the \emph{spike bus stop} via Eq: \eqref{eq:w_t_C_lim} since the $\delta^\triangle_{\text{i-th}}$ is proportional to $\bar{T}_C$. This reduction of the waiting time is only effective up to a point where the reduction in waiting time due to a lower $\bar{T}_C$ is offset by the need for more and more buses to clear the passengers at the \emph{spike bus stop}, through extra terms $\delta^\triangle_{\text{i-th}}$ in Eq. \eqref{eq:w_t_C}.
Fixing the maximum capacity $c$, synchronised bunched buses outperform perfectly staggered buses for values of $P$ high enough.

Another comparison point pertains to the applicability of the three different methods in the real world. Bunched buses are stable \cite{Chew2019, Saw2019} in the sense that perturbations in the system and in the initial conditions do not influence the long term configuration as the buses will tend to bunch. Perfectly staggered buses, on the other hand, require active dynamic intervention to keep the headway between buses constant. Such intervention can be through holding \cite{Xuan2011a,Chen2015, Daganzo2009,Saw2019intelligent,Chew2019,Abkowitz1984,rossetti1998,Hickman2001,Fu2002,cats2011,bartholdi2012,Moreira-Matias2016,Quek2020,Saw2019,Wang2020,Eberlein2001,bin2006,Mukai2008,Cortes2010,Gershenson2011,Ibarra-Rojas2015, Chen2016,Wang2018,Alesiani2018,Menda2019}, limiting boarding \cite{Delgado2009,Delgado2012,Zhao2016,Sun2018,Saw2019no,Saw2020} or stop-skipping \cite{Li1991,Eberlein1995,Fu2003,Sun2005,Cortes2010,Liu2013}. Dynamic control requires specific infrastructure and it can be challenging to implement as instructions need to be provided in real-time to bus drivers, hence a technique that involves bunched buses has the advantages of being more robust to perturbations and, at the same time, easier to deploy. 

\subsection{Comparison with a simulation}\label{ssec:simulation}
\begin{figure}[ht]
    \centering
        \includegraphics[width=16cm]{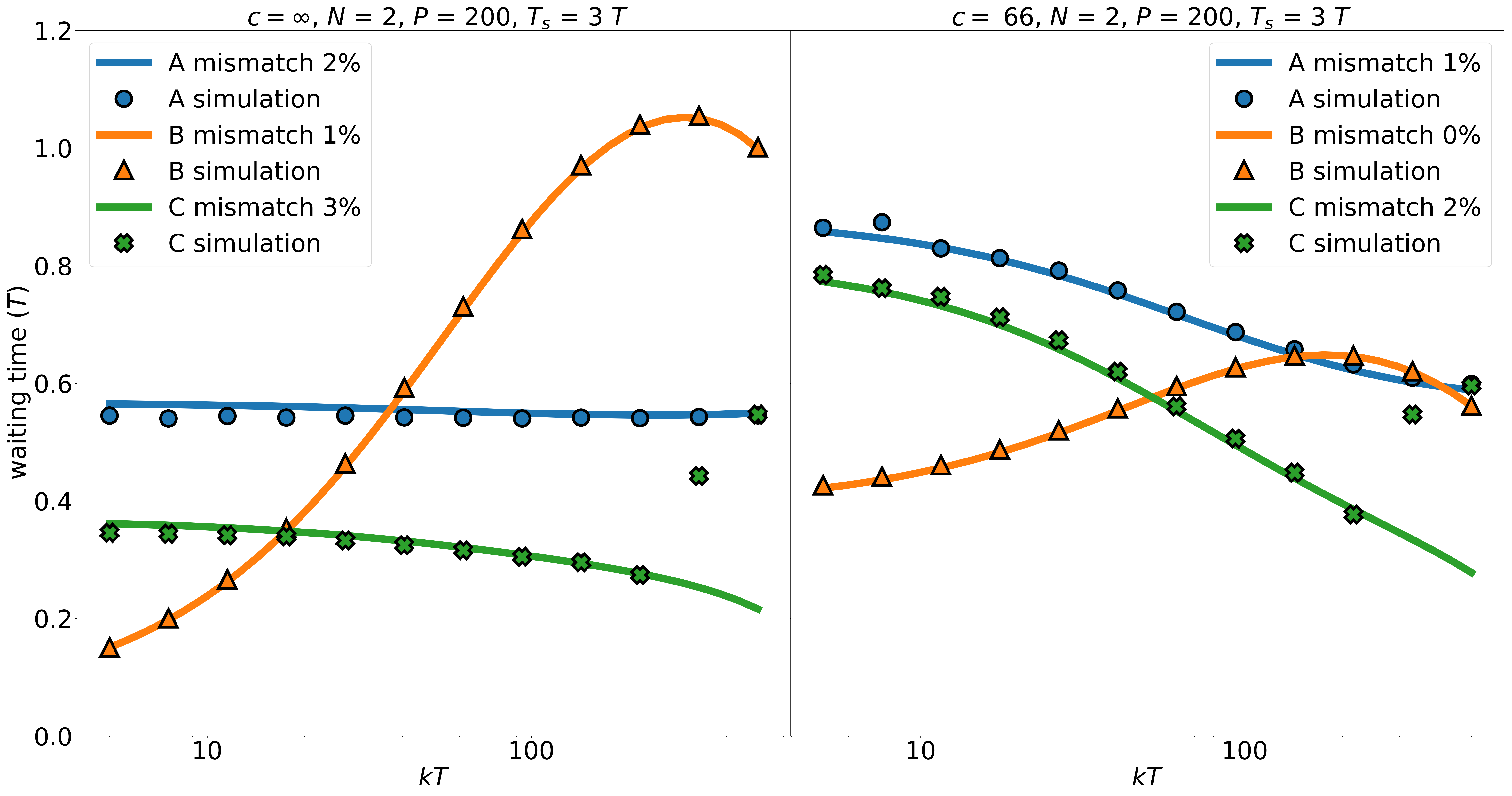}
    \caption{Comparison of the average waiting time of passengers computed with analytical formulas and with a numerical simulation. The plot on the left considers the case of unlimited capacity described in section \ref{sec:3unlimited} while the plot on the right considers a limited capacity of $c = \lfloor P / 3 \rfloor$ and uses the formulas from section \ref{sec:limited}. The circles, triangles and crosses represent the results from the simulation for the cases $A$, $B$ and $C$ respectively. Discrepancies are caused by simplified assumptions, discretisation of variables in the simulation and limited running time. The percentage of mismatch in the legend is the median of the difference between the simulation and the analytical result across the range $k T$, rounded to the closest integer. As explained in section \ref{ssec:C}, the hypothesis of perfectly staggered buses $C$ fails at high values of $k$, as we see from the plot. In this example there are two buses ($N=2)$, the period of revolution is $T = 1000$ unit times, the period of the spike is $T_s = 3 T$ and $P = 200$.}
    \label{fig:sim}
\end{figure}

To validate our analytical results, we compare the average waiting time found with our approach in sections \ref{sec:3unlimited} and \ref{sec:limited} with a time-based numerical simulation of the scenarios described. One of the challenges of the simulation is translating a result built on continuous variables in a discrete simulation, where the smallest unit is the unit of time. A convenient way to do so is to use small units of time, hence we define $T$ as 1000 units of time. The comparison with the simulation is in Fig. \ref{fig:sim}. The plot on the left considers unlimited capacity $c$ while the plot on the right compares formulas and simulation for limited $c$. The main discrepancies with the simulation are at very low $k T$, where the discretisation of time still plays a role, and at very high $k T$ for the perfectly staggered buses. This has to do with the assumption we made in section \ref{ssec:C} about the buses being perfectly staggered. In the simulation, this is implemented by holding buses at the \emph{spike bus stop} whenever the headway is not perfectly staggered, which happens after the first bus picks up passengers at the \emph{spike bus stop}. There are many choices and many ways of implementing dynamic control, one of which is described in Ref. \cite{Saw2019intelligent}, but none of them can keep the buses perfectly staggered at all times, making our assumption valid only approximately. Nonetheless, our simulation matches the waiting time predicted by the formulas within $3\%$ for most of the range of parameters explored in all three configurations analysed, both for limited and unlimited capacity.

\section{Conclusion}\label{sec:conclusion3} 
A bus loop with a \emph{spike bus stop} creates a situation where perfectly staggered buses may not be the configuration that minimises the average waiting time of passengers. If the passenger demand from the \emph{spike bus stop} is large enough, bunched buses synchronised with the spike perform better than staggered buses. The advantage is stronger when bus capacity constraints are considered. The edge of bunched buses are faster passenger boarding and higher effective capacity. When the bunched buses wait at the \emph{spike bus stop} until the spike occurs, the waiting time of the passengers arriving at the \emph{spike bus stop} is minimised, at the expenses of the waiting time of passengers arriving at the \emph{regular bus stop}.
As discussed in section \ref{sec:comparison}, bunched configurations also have the advantage of being robust to external perturbations, as well as being easier to implement as compared with staggered buses.

While our models are idealised, they take inspiration from real-world situations, such as a bus stop connected to a train station or mass rapid transit. We prove that bunched buses can outperform perfectly staggered buses in certain scenarios if the buses are synchronised with the spike in passenger demand.
The analytical results of the waiting time for the three bus configurations can be generalised to more complicated bus loops, as explained in section \ref{ssec:generalisation}.

\section*{Acknowledgements}
This work was supported by the Joint WASP/NTU Programme (Project no. M4082189).

\bibliographystyle{unsrt}
\bibliography{mybib}

\end{document}